\documentclass[floatfix, aps, prb, superscriptaddress, showkeys, reprint]{revtex4-1}

\usepackage{graphicx}               
\usepackage{bm}                     
\usepackage{amsfonts,amssymb}       
\usepackage{dcolumn}                

\newcommand*\pct{\scalebox{.9}{\%}}

\begin{document}

\title{
Anomalous thermal transport behavior in graphene-like carbon nitride (C$_3$N)
}

\author{Guangzhao~Qin}
\affiliation{Institute of Mineral Engineering, Division of Materials Science and Engineering, Faculty of Georesources and Materials Engineering, RWTH Aachen University, Aachen 52064, Germany}
\affiliation{Department of Mechanical Engineering, University of South Carolina, Columbia, SC 29208, USA}
\author{Zhenzhen~Qin}
\affiliation{International Laboratory for Quantum Functional Materials of Henan, and School of Physics and Engineering, Zhengzhou University, Zhengzhou 450001, China}
\affiliation{Aachen Institute for Advanced Study in Computational Engineering Science (AICES), RWTH Aachen University, Aachen 52062, Germany}
\author{Huimin~Wang}
\affiliation{College of Engineering and Applied Science, Nanjing University, Nanjing 210023, China}
\affiliation{Institute of Mineral Engineering, Division of Materials Science and Engineering, Faculty of Georesources and Materials Engineering, RWTH Aachen University, Aachen 52064, Germany}
\author{Jianjun~Hu}
\affiliation{Department of Computer Science and Engineering, University of South Carolina, Columbia, SC 29208, USA.}
\author{Ming~Hu}
\email{hu@sc.edu}
\affiliation{Department of Mechanical Engineering, University of South Carolina, Columbia, SC 29208, USA}

\date{\today}


\begin{abstract}
The graphene's success created a new era in materials science, especially for
two-dimensional (2D) materials.
New classes 2D carbon-based materials beyond graphene have been intensively
studied for their promising applications in nano-/opto-/spin-electronics,
catalysis, sensors, clean energy, \emph{etc}.
Very recently, the controllable large-scale synthesis of 2D single crystalline
carbon nitride (C$_3$N) was reported, which is the first and the only
crystalline, hole-free, single-layer carbon nitride.
Due to the fascinating properties, monolayer C$_3$N attracts tremendous interest
for the potential applications in various fields, where thermal transport
property is severely concerned for developing high performance devices in terms
of advanced thermal management, which, however, remain less investigated.
Here, we perform a comparative study of thermal transport between monolayer
C$_3$N and the parent graphene, by solving the phonon Boltzmann transport
equation (BTE) based on first-principles calculations.
The thermal conductivity ($\kappa$) of C$_3$N shows an anomalous temperature
dependence, which is totally different from that for common crystalline
materials and deviates largely from the well-known $\kappa\sim 1/T$
relationship.
Consequently, the $\kappa$ of C$_3$N at high temperatures is larger than the
expected value following the common trend of $\kappa \sim 1/T$.
Moreover, the $\kappa$ of C$_3$N is found in surprise to be enlarged by applying
bilateral tensile strain, despite its similar planar honeycomb structure as
graphene, whose $\kappa$ is reduced upon stretching.
Thus, it would benefit the applications of C$_3$N in nano- and opto-electronics
in terms of efficient heat dissipation, considering the unexpectedly large
$\kappa$ at high working temperature or under realistic conditions where
residual strain usually exists after fabrication or synthesis.
The underlying mechanism is revealed by providing direct evidence for the
interaction between lone-pair N-$s$ electrons and bonding electrons from C atoms
in C$_3$N based on the analysis of orbital-projected electronic structures and
electron localization function (ELF).
Our study not only make a comprehensive investigation of the thermal transport
in graphene-like C$_3$N, but also reveals the physical origins for its anomalous
properties, which deepens the understanding of phonon transport in 2D materials
and would also have great impact on future research in micro-/nanoscale thermal
transport such as materials design with targeted thermal transport properties.
\end{abstract}

\maketitle

\section{Introduction}

\begin{figure}[tb]
    \centering
    \includegraphics[width=0.95\linewidth]{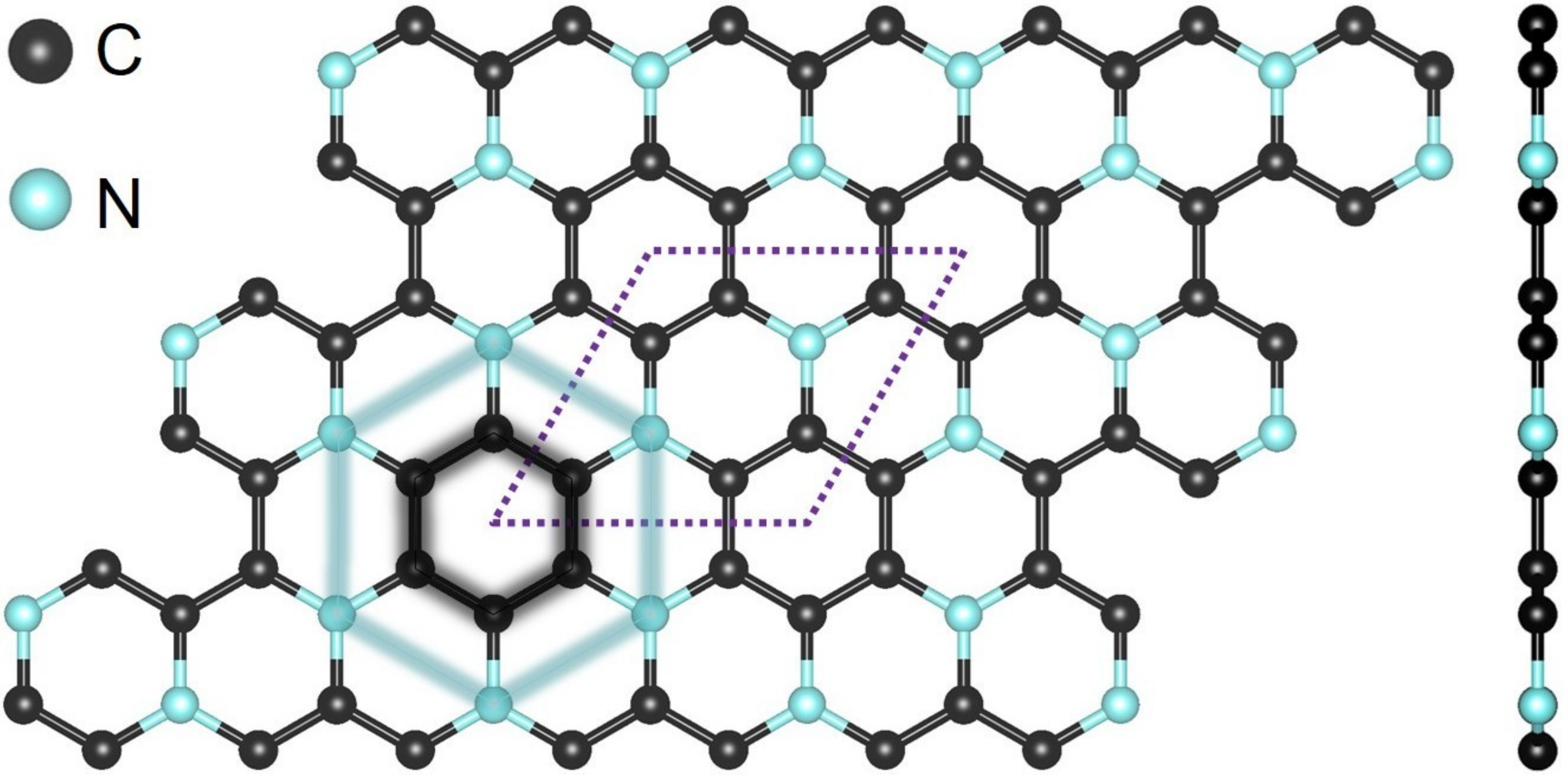}
\caption{\label{fig:structure}
(Color online)
The top (left) and side (right) view of geometry structure of monolayer C$_3$N,
showing graphene-like planar morphology.
The primitive cell (8 atoms) is marked with the dot diamond.
Colored cyan and black shadow lines indicate the hexagonal structures formed by
atoms C and N, respectively.
}
\end{figure}

Graphene, an atomically thin two-dimensional (2D) material with honeycomb
lattice structure, exhibits numerous striking physical properties, and can, in
principle, be considered as an elementary building block for various carbon
allotropes.
Ever since the recent developments in 2004, the field of graphene research took
off rapidly.
These developments in the science of graphene prompted an unprecedented surge of
activity and demonstration of new physical phenomena with novel
applications\cite{Balandin2012266, NatMater.2007.6.3.183-191}, such as
nanoelectronics, energy storage and conversion, medicine, catalysis, sensors,
\emph{etc}.
Graphene has lots of excellent properties such as the surprisingly large
room-temperature electron mobility, high strength and flexibility, and record
high thermal conductivity\cite{RevModPhys.81.109, Zhang2015,
Appl.Phys.Lett..2012.101.11.111904, Phys.Rev.B.2015.91.3.035416}.
However, attempts to utilize graphene for practical applications are faced with
some limitations, especially the poor on-off current ratio
($I_\mathrm{ON}/I_\mathrm{OFF}<100$) in graphene-based devices due to the
gapless nature of graphene\cite{Nature.2011.479.7373.338-344}.
Thus, the substitutions of carbon (C) atom in graphene with heteroatoms were
stimulated for the extension of graphene family to other 2D layered crystalline
materials\cite{Nanoscale.2015.7.11.4598-4810}.
Among these, monolayer hexagonal boron nitride ($h$-BN) with a wide band gap
($\sim$5.0-6.0\,eV) offers an alternative solution beyond the gapless graphene,
which establishes the key role of 2D nitrides in advancing the development of
next generation nano-electronics\cite{MaterialsResearchLetters.2013.1.4.200-206}.
To benefit from carbon-based nanomaterials at the same time, partially
substituting C atoms in graphene with N is a plausible approach to the formation
of graphene-like 2D carbon nitrides.
In the past years, different N/C ratios have been realized\cite{Science.2015.347.6225.970,NatMater.2009.8.1.76-80,Nat.Commun..2015.6..6486}.
For example, 2D crystalline layered C$_3$N$_4$ and C$_2$N-$h2D$ are
semiconductors with direct bandgaps of 2.76 and 1.96\,eV, respectively, with
potential applications in nanoelectronics, photo-catalysis, solar power
generation, \emph{etc}.
However, large number of holes exist in the crystalline structures due to the
large N/C ratios.
Very recently, Yang \emph{et al.}\cite{AdvancedMaterials.2017.29.16.1605625}
reported the controllable large-scale (up to hundreds of micrometer) synthesis
of 2D single crystalline carbon nitride (C$_3$N) sheet, which is the first and
the only crystalline, hole-free, single-layer carbon nitride, showing
graphene-like morphology (Fig.~\ref{fig:structure}).

Monolayer C$_3$N possesses the graphene-like planar honeycomb structure with a
homogeneous distribution of C and N atoms, both of which show the
D$_{6h}$-symmetry\cite{PNAS.2016.113.27.7414-7419}.
Despite various 2D carbon-based materials, C$_3$N is the only one possessing
indirect bandgap, which is 0.39\,eV as verified both experimentally and
theoretically and can be tuned to cover the entire visible
range\cite{AdvancedMaterials.2017.29.16.1605625}.
Back-gated field-effect transistors (FET) made of monolayer C$_3$N display an
on-off current ratio reaching 5.5$\times 10^{10}$, which is much larger than
those of graphene ($< 10^2$), C$_2$N ($10^7$), and phosphorene ($\sim 10^3$)\cite{Nat.Commun..2015.6..6486}.
Moreover, C$_3$N hydrogenated with a sufficient amount of hydrogen shows
spontaneous magnetism (ferromagnetic, FM) at temperatures lower than 96\,K.
Considering almost all the novel applications of C$_3$N in nanoelectronics are
inevitably involved with heat dissipation, the thermal transport properties are
of great interest for developing high performance C$_3$N-based devices in terms
of efficient thermal management.

Efficient regulation of heat transfer plays a key role in the high-performance
thermal management of nanoetechnologies.
The heat conduction in semiconductors is mainly carried out by phonon transport
\cite{NatMater.2011.10.8.569-581,gu2015phonon,Balandin2012266}.
Therefore, the fundamental understanding of phonon transport is of great
significance for the effective control of heat flow, and it is a thermophysical
problem that has great practical significance related to energy technology, such
as electrical cooling, thermoelectric technology, phase change storage,
\cite{Balandin2012266,AppliedPhysicsReviews.2014.1.1.011305}
thermal devices (diodes, transistors, logic gates),
\cite{Phys.Rev.Lett..2004.93.18.184301}
\emph{etc}.
In the past few decades, lots of studies have focused on the effective
regulation of heat transport by nanostructuring
\cite{NanoLett..2011.11.2.618-623,Science.2007.315.5810.351,NatMater.2017.16.1.83-88}.
Besides, pores are introduced into the physicochemical treatment (hydrogenation,
oxidation, \emph{etc})\cite{NanoLett..2014.14.4.1734-1738}.
In addition, effective thermal modulation can also be achieved by external
electric/magnetic field\cite{Nanoscale.2017.9.21.7227-7234} and strain
engineering\cite{Phys.Rev.B.2016.93.7.075404,Phys.Rev.B.2010.81.24.245318,NanoLett..2016.16.6.3831-3842}.
Due to the robust reliability and strong flexibility, strain engineering has
become one of the most promising and effective ways to achieve continuously
adjustable heat transport.
Moreover, the actual case of many systems and devices typically contain residual
strain after fabrication\cite{TsinghuaScienceandTechnology.Feb.2009.14.1.62-67}.
Therefore, the study of strain engineering on the regulation of thermal
conductivity has a very important practical significance.
However, previous studies have mainly focused on how to adjust thermal
conductivity through mechanical strain, and there is still a lot of unclear
understanding of the essential origin of its regulatory effects
\cite{TheJournalofChemicalPhysics.2006.125.16.164513,Phys.Rev.B.2014.90.23.235201,NanoLett..2016.16.6.3831-3842,Nanotechnology.2016.27.26.265706,Phys.Rev.B.2016.93.7.075404,PhysRevB.87.195417}.
These in-depth understanding benefits more effective and accurate thermal
conductivity regulation, which would have a far-reaching guiding role.
Thus, the modulation of the thermal transport properties of monolayer C$_3$N by
mechanical strain could be practically meaningful, and the origin of the
underlying mechanism would deepen our understanding of phonon transport in 2D
materials and have great impact on future research in materials design with
targeted thermal transport properties.

In this paper, by solving the phonon Boltzmann transport equation (BTE) based on
first-principles calculations, we perform a comparative study of phonon
transport between monolayer C$_3$N and graphene.
Besides the anomalous temperature dependence of $\kappa$ of C$_3$N, it is very
intriguing to find that the $\kappa$ of C$_3$N is more than one order of
magnitude lower than graphene, considering the similar structures and the only
difference of substituting 1/4 C with N atoms in C$_3$N compared to graphene.
By deeply analyzing the orbital projected electronic structure, we establish a
microscopic picture of the lone-pair electrons driving strong phonon
anharmonicity.
We show that nonlinear restoring forces arise from the interactions between
lone-pair electrons around N atoms and bonding electrons from adjacent atoms
(C), leading to strong phonon anharmonicity and low $\kappa$.
Furthermore, the $\kappa$ of C$_3$N is unexpectedly enlarged by applying
bilateral tensile strain despite the planar honeycomb structure of C$_3$N
(similar to graphene, with no buckling or puckering), which is in sharp contrast
to the strain induced $\kappa$ reduction in graphene.
The opposite response of $\kappa$ to mechanical strain between C$_3$N and
graphene further supports the established microscopic picture of the lone-pair
electrons driving strong phonon anharmonicity.

\begin{figure*}[tb]
    \centering
    \includegraphics[width=0.95\linewidth]{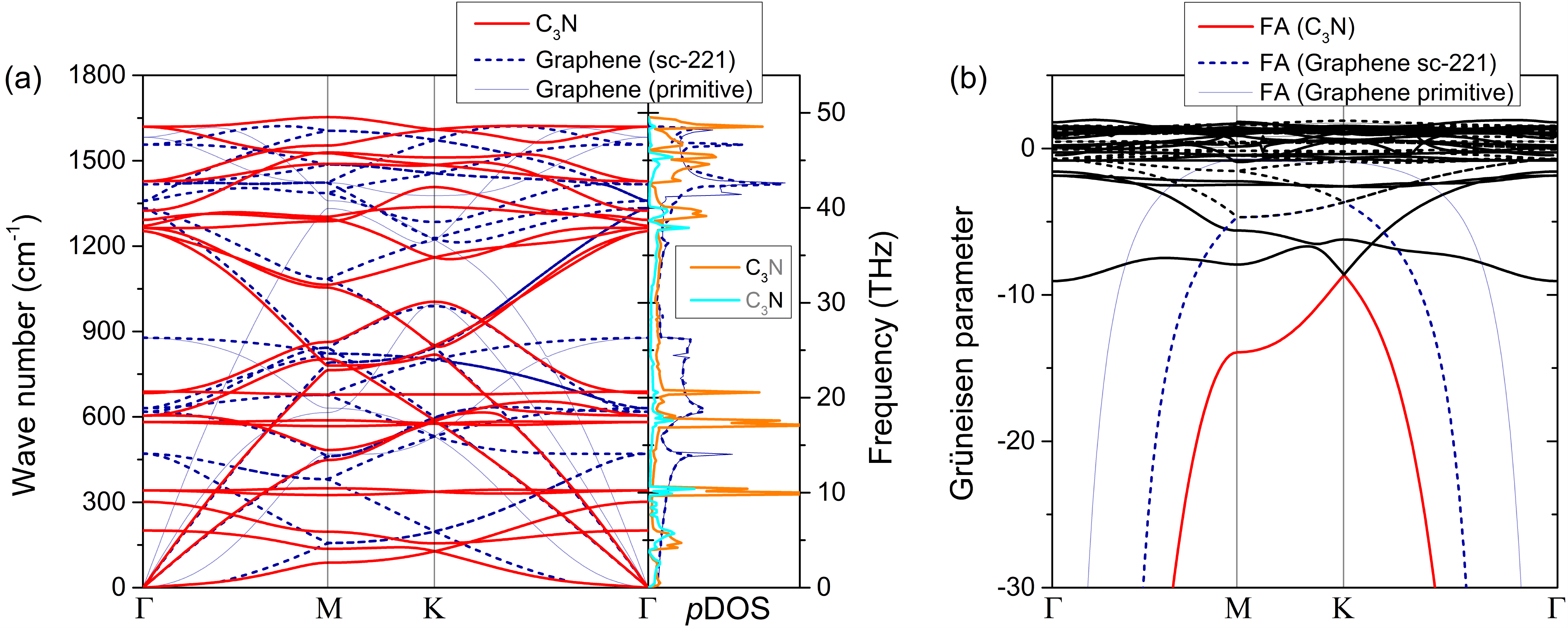}
\caption{\label{fig:dispersion}
(Color online)
Comparison of (a) phonon dispersions, partial density of states ($p$DOS), and
(b) Gr\"uneisen parameters between C$_3$N and graphene.
For the purpose of comparison, the supercell of $2\times 2\times 1$ (8 atoms,
the same as C$_3$N) is used as the unit cell of graphene.
Also plotted are the results of graphene with primitive cell (2 atoms).
C$_3$N possesses softened FA phonon branch and the corresponding strong phonon
anharmonicity.
}
\end{figure*}

\section{Methods}

All the first-principles calculations are performed in the framework of density
functional theory (DFT) using the projector augmented wave (PAW) method
\cite{PhysRevB.59.1758} as implemented in the Vienna \emph{ab initio} simulation
package (\texttt{\textsc{vasp}})\cite{PhysRevB.54.11169}.
The Perdew-Burke-Ernzerhof (PBE) of generalized gradient approximation (GGA)
revised for solids (PBEsol)\cite{Phys.Rev.Lett..2008.100.13.136406} is chosen as
the exchange-correlation functional\cite{CMS.2018.151.0927}.
The kinetic energy cutoff of wave functions is set as 1000\,eV and a
Monkhorst-Pack \cite{PhysRevB.13.5188} $k$-mesh of $31\times 31\times 1$ is used
to sample the Brillouin Zone (BZ) with the energy convergence threshold of
$10^{-6}$\,eV, where the $k$ is the electronic wavevector spanned by the
reciprocal lattice vectors.
A large enough vacuum spacing of 20\,\AA\ is used along the \emph{out-of-plane}
direction based on the convergence test.
The applied biaxial strain is defined as $(l-l_0)/l_0$, where $l$ is the lattice
constant under strains and $l_0$ corresponds to the original value with no
strain applied.
All geometries are fully optimized based on the primitive cell containing 8
atoms (6 C and 2 N as shown in Fig.~\ref{fig:structure}) to get the stable
configuration with globally minimized energy, until the maximal Hellmann-Feynman
force is smaller than $10^{-8}$\,eV/\AA.
For the consistent calculations of interatomic force constants (IFCs) using the
finite displacement difference method, $3\times 3\times 1$ supercell containing
72 atoms is constructed and the Monkhorst-Pack $k$-mesh of $2\times 2\times 1$
is used to sample the BZ, which can accurately describe the system based on the
convergence test
\cite{Phys.Rev.B.2017.95.19.195416, Nanoscale.2017.9.12.4295-4309}.
The space group symmetry properties are used to reduce the computational cost
and the numerical noise of the forces \cite{phonopy}.
The translational and rotational invariance of IFCs are enforced using the
Lagrange multiplier method
\cite{PhysRevB.77.144112, PhysRevB.86.174307}.
The Born effective charges ($Z^*$) and dielectric constants ($\epsilon$) are
obtained based on the density functional perturbation theory (DFPT), which are
added to the dynamical matrix as a correction to take long-range electrostatic
interaction into account.
Based on kinetic theory, $\kappa$ can be expressed as\cite{Phys.Rev.B.2005.72.1.014308}
\begin{equation}
\label{eq:kappa}
\kappa_{\alpha} =
    \sum_{\vec{q}p}C_\mathrm{V} (\vec{q},p)
    v_\alpha(\vec{q},p)^2
    \tau(\vec{q},p)
\ ,
\end{equation}
where $C_\mathrm{V}$ is the volumetric specific heat capacity of phonon
following the Bose-Einstein statistics, $\vec{v}_\alpha(\vec{q},p)$ is the
$\alpha(=x,y,z)$ component of group velocity of phonon mode with wave vector
$\vec{q}$ and polarization $p$, and $\tau$ is the relaxation time (phonon
lifetime).
The $\kappa$ is obtained by iteratively solving the phonon BTE with the ShengBTE
package, which is equivalent to the solution of relaxation time approximation
(RTA) if the iteration stops at the first step
\cite{Li20141747, PhysRevB.86.174307}.
The convergence test of $\kappa$ with respect to the cutoff distance and
$Q$-grid are fully conducted\cite{npjCM.2018.4.1.3}, based on which the cutoff
distance is chosen as 7.36\,\AA\ (13th nearest neighbors) for the 3rd IFCs
calculations and the $Q$-grid is chosen as $50\times 50\times 1$ for the
$\kappa$ calculations.
The $\kappa$ of graphene is calculated to be 3094.98\,W/mK by using the
iterative method.
The good agreement with previous results confirms the reliability of our
calculations\cite{Balandin2012266}.

\section{Structure and phonon dispersion}

Up to now, monolayer C$_3$N is the only crystalline, hole-free, single-layer
carbon nitride\cite{AdvancedMaterials.2017.29.16.1605625}, which shows
graphene-like morphology (Fig.~\ref{fig:structure}).
C$_3$N possesses a planar honeycomb structure, different from the buckled
silicene and the puckered phosphorene, with the primitive cell containing 8
atoms (6 C and 2 N), in which both C and N atoms show a D$_{6h}$-symmetry.
The optimized lattice constant of C$_3$N is 4.86\,\AA.
The C-C and C-N bond lengths in C$_3$N are calculated to be slightly different
(1.40326 and 1.40288\,\AA, respectively).
Based on the optimized structure, we calculate the phonon dispersions and
partial density of states ($p$DOS) of C$_3$N [Fig.~\ref{fig:dispersion}(a)].
No imaginary frequency is observed, indicating the thermodynamical stability of
monolayer C$_3$N.
The results of graphene are also plotted for comparison, which are calculated
with the supercell of $2\times 2\times 1$ (i.e.\ 8 atoms, the same as C$_3$N)
used as the unit cell.
We also plot the results of graphene with primitive cell (2 atoms) to give a
hint.
The flexural acoustic (FA) phonon branches ($z$-direction vibration) of both
C$_3$N and graphene show the quadratic behavior, which is the typical feature of
2D materials\cite{RevModPhys.81.109} and is also observed in silicene and
phosphorene\cite{PhysRevB.89.054310, Phys.Chem.Chem.Phys..2015.17.7.4854-4858,
Phys.Rev.B.2016.94.16.165445}.
The phonon dispersions of C$_3$N and graphene are highly consistent, especially
for the longitudinal and transverse acoustic phonon branches.
However, the FA phonon branch of C$_3$N is significantly softened (lower
frequency and smaller group velocity) compared to graphene, which suggests
possibly strong phonon anharmonicity.

Usually, the strength of the phonon anharmonicity can be quantified by the
Gr\"uneisen parameter, which characterizes the relationship between phonon
frequency and crystal volume change.
Thus, we further calculate the Gr\"uneisen parameters of C$_3$N and graphene
[Fig.~\ref{fig:dispersion}(b)] to quantitatively assess the phonon
anharmonicity.
The magnitude of the Gr\"uneisen parameters of C$_3$N are very large compared to
graphene, especially for the FA phonon branch, confirming stronger phonon
anharmonicity in C$_3$N.
Considering the similar structures and the only difference of substituting 1/4 C
with N atoms in C$_3$N compared to graphene, it is very intriguing to find the
significantly softened FA phonon branch and strong phonon anharmonicity in
C$_3$N.
We will show later the electronic origin of the orbital driven strong phonon
anharmonicity and the anomalous large magnitude of Gr\"uneisen parameter in
C$_3$N.
As it is known that strong phonon anharmonicity can give rise to low $\kappa$ in
ordered crystal structures, we would further study the thermal transport
properties of C$_3$N.

\begin{figure}[tb]
    \centering
    \includegraphics[width=0.95\linewidth]{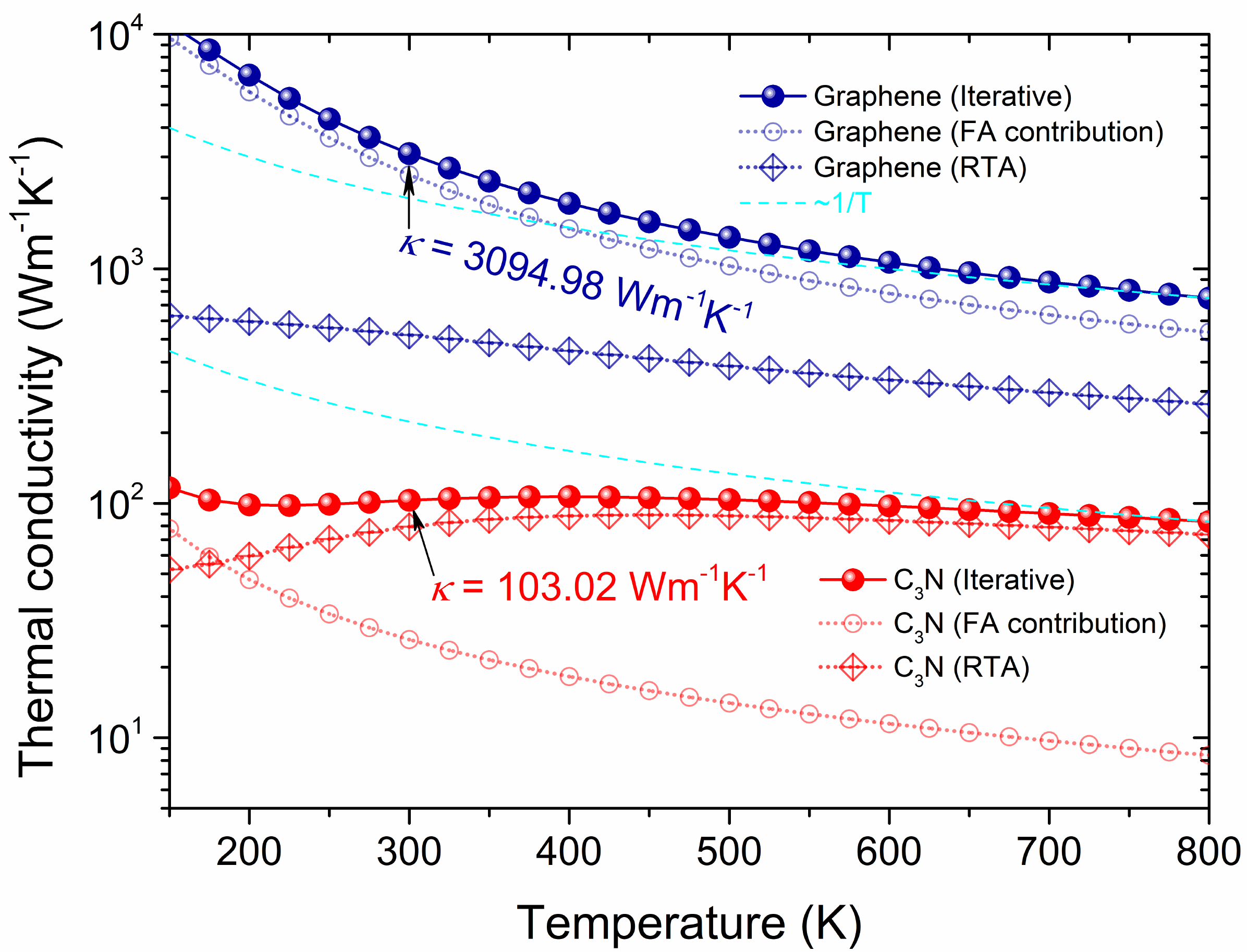}
\caption{\label{fig:kappa-T}
(Color online)
The temperature dependent $\kappa$ of C$_3$N and graphene calculated using
iterative and RTA methods.
The $\kappa$ at 300\,K is indicated with arrows.
The contributions from FA phonon branch and the $\kappa \sim 1/T$ relations are
also plotted for comparison.
The $\kappa$ of C$_3$N is more than one order of magnitude lower than that of
graphene and shows an almost temperature independence from 200 to 800\,K.
}
\end{figure}

\section{Thermal transport properties}

Fig.~\ref{fig:kappa-T} shows the temperature dependent $\kappa$ of C$_3$N
obtained by iteratively solving the phonon BTE together with the results from
RTA method, in comparison with graphene.
The room temperature $\kappa$ of C$_3$N is 103.02\,W/mK, which is more than one
order of magnitude lower than that of graphene (3094.98\,W/mK).
Note that Mortazavi\cite{Carbon.2017.118..25-34} reported a rather high $\kappa$
of C$_3$N (815\,W/mK) based on the classical non-equilibrium molecular dynamics
(MD) simulations, where the optimized Tersoff and original Tersoff potentials
are used to describe the C-C and C-N interatomic interactions, respectively.
Considering that classical MD simulation suffers from the accuracy of the
empirical potential used\cite{PhysRevB.89.054310}, the discrepancy of the
$\kappa$ between ours and Mortazavi \emph{et al.}'s could be attributed to the
different computational methods employed (first-principles \emph{vs.}\ empirical
potential).
For example, the C-C and C-N bond lengths in C$_3$N are calculated to be 1.44
and 1.43\,\AA\ in classical MD simulations, respectively\cite{Carbon.2017.118..25-34},
which also differ quite largely from the results by first-principles (1.40326 and
1.40288\,\AA, respectively).
The $\kappa$ of C$_3$N obtained from the RTA method is very close to the
accurate value obtained by iteratively solving the phonon BTE, suggesting small
proportion of N-process and weak phonon hydrodynamics in C$_3$N.
Previous studies showed that the phonon hydrodynamics due to momentum conserving
processes is responsible for the ultra-high $\kappa$ of graphene
\cite{NatCommun.2015..6.6400,NatCommun.2015.6..6290,PhysicsReports.2015.595..1-44}.
Thus, the weak effect of phonon hydrodynamics in C$_3$N would be responsible for
the significantly lower $\kappa$ of C$_3$N compared to graphene.
The FA contribution to the room temperature $\kappa$ of C$_3$N is 25.4\pct,
which contributes the most compared to other phonon branches but is not large
enough to dominate the phonon transport as the case for graphene (81.2\pct).
More strikingly, as shown in Fig.~\ref{fig:kappa-T}, the $\kappa$ of C$_3$N
shows an anomalous temperature dependence, which deviates largely from the
well-known $\kappa\sim 1/T$ relation, being quite different from common cases of
crystalline materials
\cite{Appl.Phys.Lett..2016.109.24.242103,
Phys.Chem.Chem.Phys..2015.17.7.4854-4858, Nanoscale.2016.8.21.11306-11319,
Phys.Rev.B.2016.94.16.165445, Phys.Rev.B.2016.93.7.075404,
Phys.Rev.B.2011.84.15.155421, ComputationalMaterialsScience.2015.110..115-120,
Phys.Rev.Lett..2012.109.9.095901, NatPhys.2015.11.12.1063-1069,
apl.10510101907.1.4895770, Sci.Rep..2016.6..19830,
Phys.Rev.B.2013.87.16.165201}.
In fact, the $\kappa$ of C$_3$N has no considerable change over a large
temperature range (200-800\,K) due to the anomalous temperature dependence.
Consequently, the $\kappa$ of C$_3$N at high temperature is larger than the
expected value following the common $\kappa \sim 1/T$ trend, which would largely
benefit its applications in nano- and opto-electronics in terms of efficient
heat dissipation.

\begin{figure}[tb]
    \centering
    \includegraphics[width=0.99\linewidth]{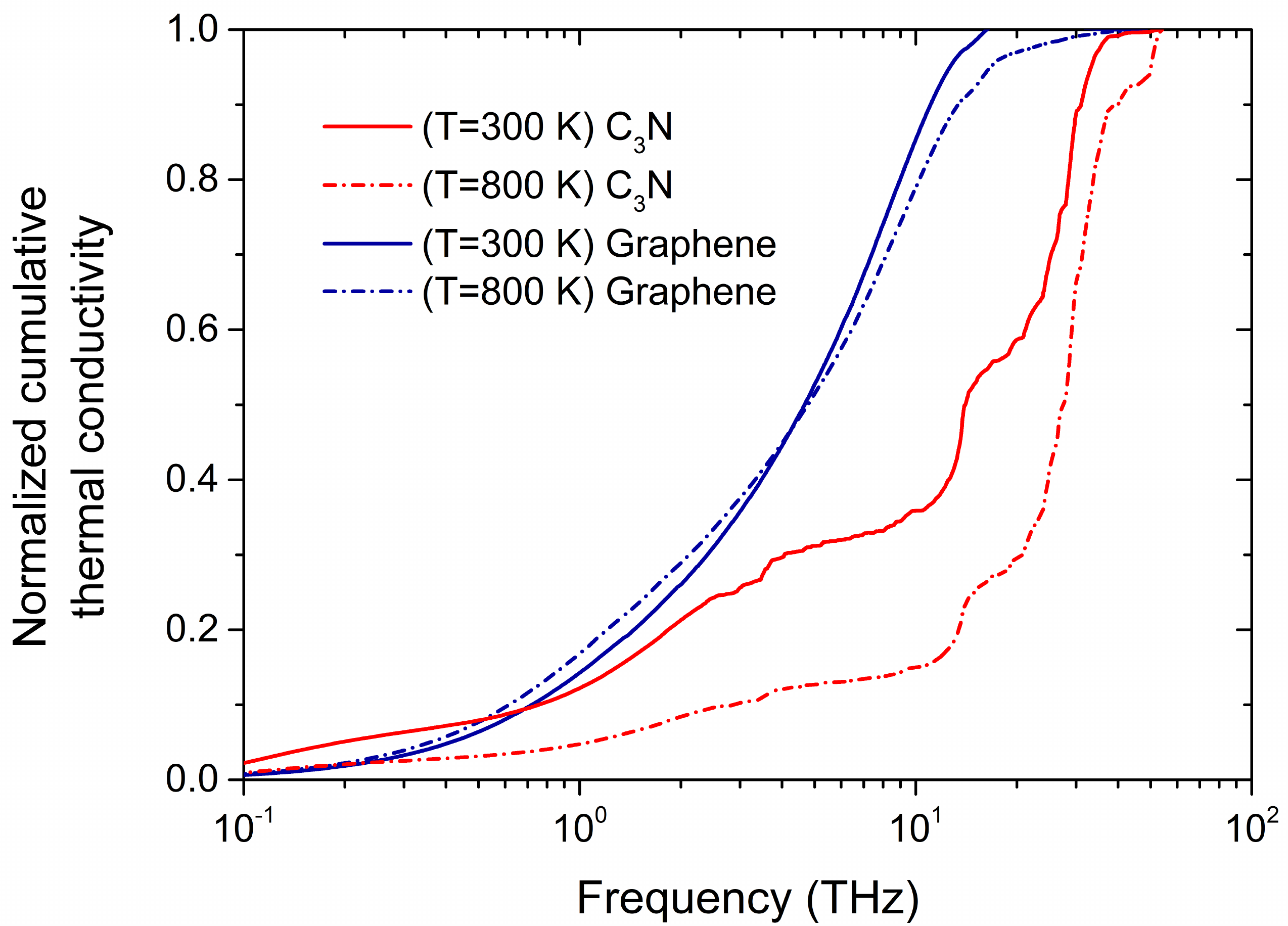}
\caption{\label{fig:cumulative-freq}
(Color online)
Comparison of the normalized cumulative $\kappa$ with respect to frequency
between C$_3$N and graphene at 300 and 800\,K.
The relatively large contribution of high-frequency phonon modes in C$_3$N is
the reason for the anomalous temperature dependence of $\kappa$.
}
\end{figure}

\begin{figure*}[tb]
    \centering
    \includegraphics[width=0.99\linewidth]{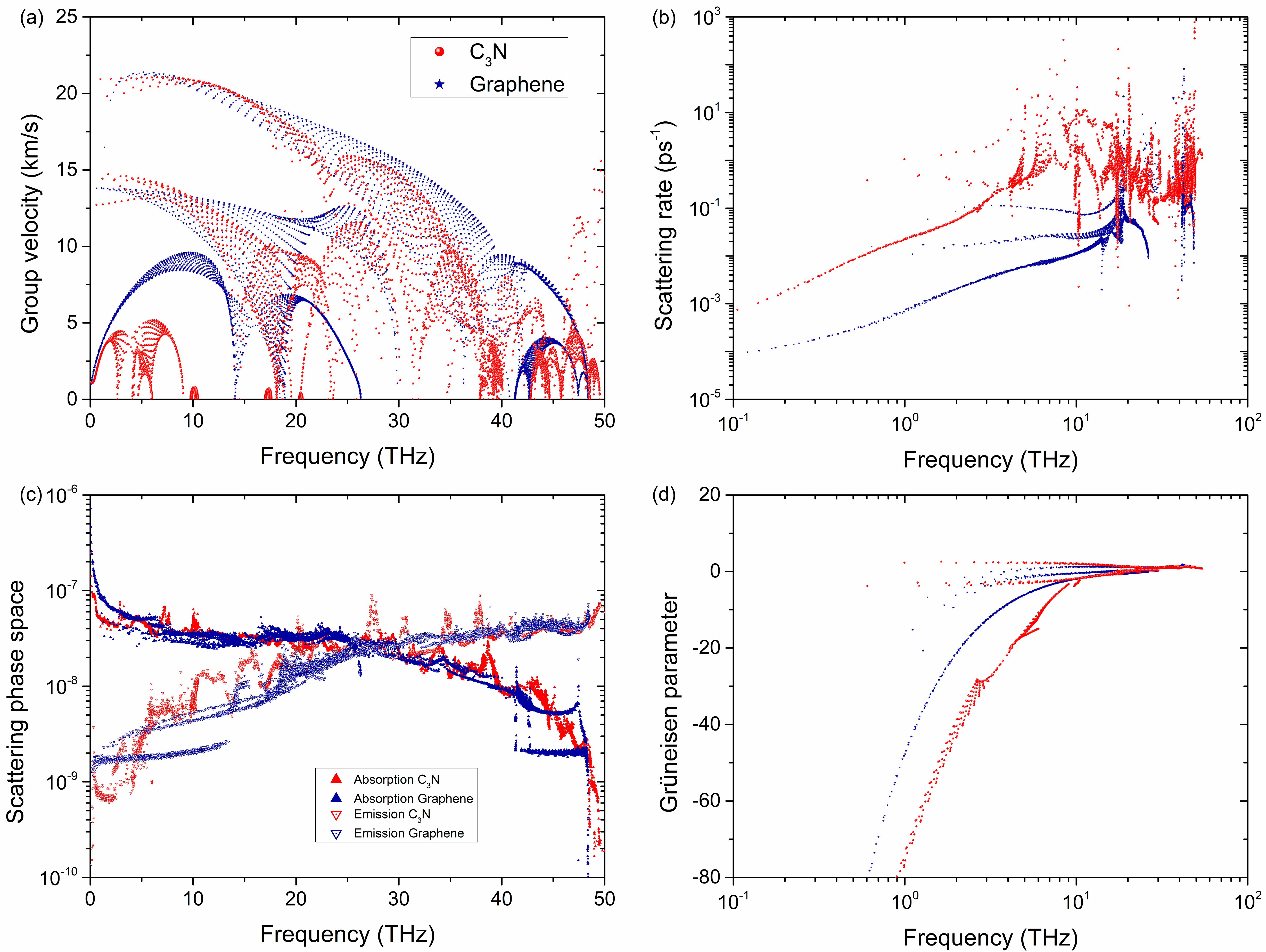}
\caption{\label{fig:mode}
(Color online)
Comparison of mode level (a) phonon group velocity, (b) scattering rate, (c)
scattering phase space (absorption and emission process), and (d) Gr\"uneisen
parameter between C$_3$N and graphene.
The significantly lower $\kappa$ of C$_3$N than that of graphene lies in the
larger scattering rate, which is caused by the larger phonon anharmonicity
instead of the scattering phase space.
}
\end{figure*}

\subsection{Anomalous temperature dependence}

To understand the underlying mechanism of the anomalous temperature dependence
of $\kappa$ of C$_3$N, we compare the frequency accumulated $\kappa$ of C$_3$N
and graphene in Fig.~\ref{fig:cumulative-freq}.
For graphene, the main contribution to $\kappa$ is from low-frequency acoustic
phonon modes at both low and high temperatures.
However, the situation in C$_3$N is quite different from that in graphene.
High-frequency optical phonon modes contribute largely to the $\kappa$ of
C$_3$N, especially when the temperature becomes high.
It is well-known that the lower the frequency, the lower the required temperature
for thermal activation\cite{Phys.Rev.B.2017.95.19.195416}.
The low frequency phonon modes can be much more easily thermally activated and
already saturate at a lower temperature than the concerned temperatures.
Whereas, high temperature is required for the thermal activation of high
frequency phonon modes, which just begin to be thermally activated around the
concerned temperatures.
As revealed in Eq.~(\ref{eq:kappa}), the temperature dependence of $\kappa$ is
the competition between heat capacity and phonon mean free path (MFP) that
follows the $\sim 1/T$ law.
Thus, the temperature dependence of $\kappa$ contribution from low-frequency
phonon modes is dominated by the phonon MFP and decreases with temperature
increasing with the relation of $\sim 1/T$.
Whereas for high-frequency phonon modes, the variation is dominated by the heat
capacity due to its fast increase with the thermal activation and the
$\kappa$ contribution increases quickly with temperature increasing, making the
temperature dependence of $\kappa$ deviates largely from the well-known
$\kappa\sim 1/T$ relation.
Hence, the relatively large contribution of high-frequency phonon modes in
C$_3$N (Fig.~\ref{fig:cumulative-freq}) is the direct reason for the anomalous
temperature dependence of $\kappa$.

\begin{figure}[tb]
    \centering
    \includegraphics[width=0.99\linewidth]{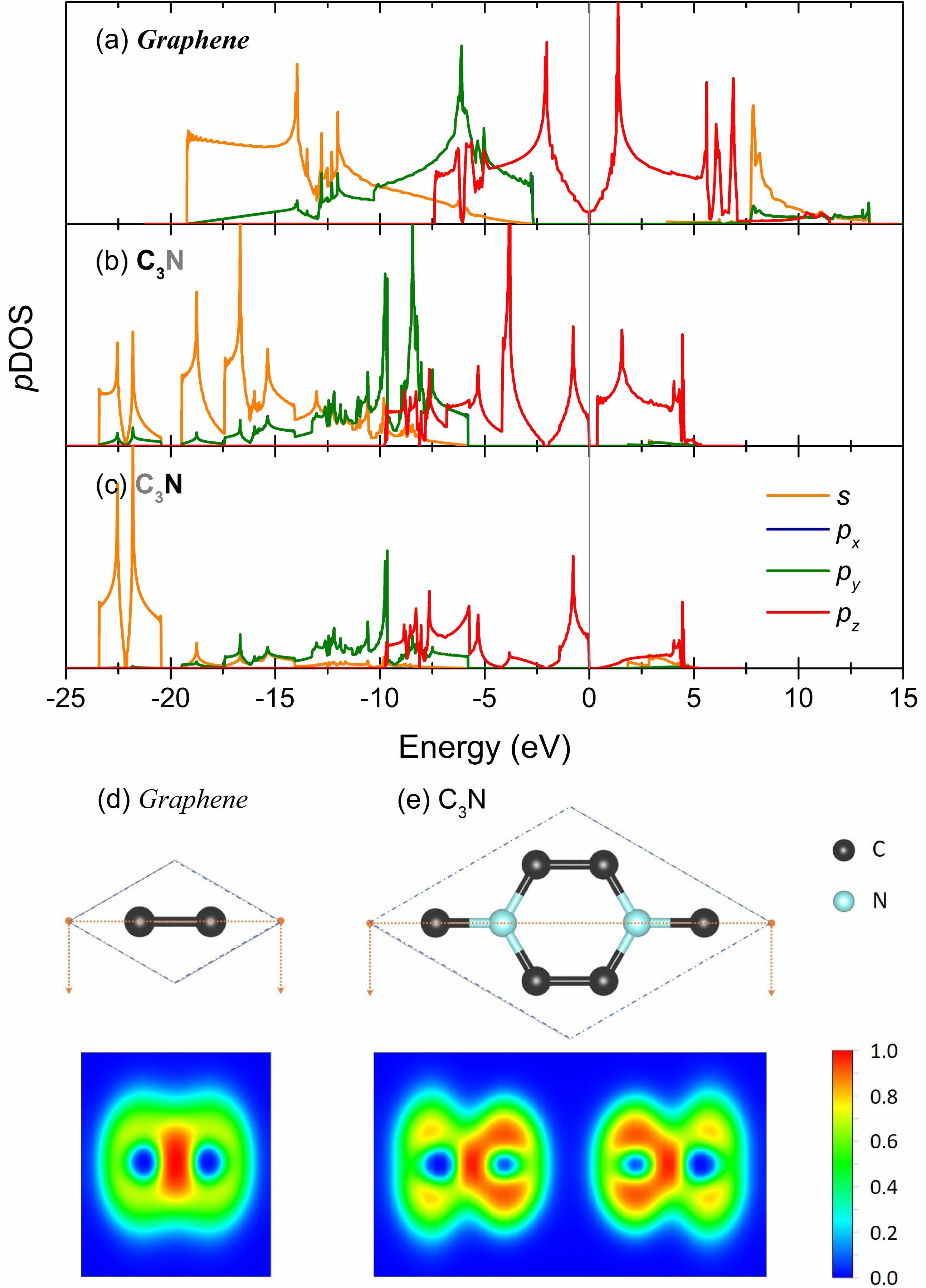}
\caption{\label{fig:pDOS}
(Color online)
The $s$, $p_x$, $p_y$, and $p_z$ orbital projected electronic density of states
($p$DOS) of (a) graphene and C$_3$N with the contributions from atoms (b) C and
(c) N decoupled.
Note that the in-plane $p_x$ and $p_y$ orbitals overlap with each other.
(d,e) Side view of the electron localization function (ELF) of (d) graphene and (e) C$_3$N.
The atomic representations are top view and the ELF is taken along the orange
dotted line.
The non-bonding lone-pair $s$ electrons arise around N atoms in C$_3$N, leading
to strong phonon anharmonicity by interacting with the covalently bonding
electrons of adjacent atoms (C).
}
\end{figure}

In addition to C$_3$N, the anomalous temperature dependence of $\kappa$
was also found in monolayer gallium nitride
(GaN)\cite{Phys.Rev.B.2017.95.19.195416} and zinc oxide
(ZnO)\cite{Phys.Chem.Chem.Phys..2017.Huimin.ZnO}, where the high-frequency
phonon modes also contribute largely to the $\kappa$.
It was analyzed
that\cite{Phys.Rev.B.2017.95.19.195416,Phys.Chem.Chem.Phys..2017.Huimin.ZnO} in
monolayer GaN and ZnO, the large contribution from high-frequency phonon modes
is due to the enhanced phonon group velocity and the relatively large phonon
lifetime, which are further traced back to the strongly polarized bond due to
the electronegativity and the huge phonon bandgap in the phonon dispersion due
to the difference in atom mass.
Here, although the electronegativity and the difference in atom mass is not
large for C$_3$N, the group velocity of LO phonon branch is still enhanced
[Fig.~\ref{fig:mode}(a)] due to the LO-TO splitting caused by the polarization
in C-N bond [Inset of Fig.~\ref{fig:pDOS}(c)].
Besides, the lifetime of high-frequency phonon modes in C$_3$N is relatively
large due to the weakened phonon-phonon scattering [Fig.~\ref{fig:mode}(b)] (the
scattering rate is comparable to graphene while that of low-frequency phonon
modes is much larger) caused by the phonon bunching and flattening
[Fig.~\ref{fig:dispersion}(a)].
Thus, the high-frequency phonon modes contribute largely to the $\kappa$ of
C$_3$N (Fig.~\ref{fig:cumulative-freq}), which results in the anomalous
temperature dependence of $\kappa$.
It is worth pointing out that, a large difference in atom mass and consequently
a huge bandgap in the phonon dispersion as analyzed in previous
study\cite{Phys.Rev.B.2017.95.19.195416} are not necessary for the anomalous
temperature dependence of $\kappa$, while phonon bunching and flattening can
also have the similar effect.

\subsection{Strong phonon anharmonicity and low thermal conductivity}

The room temperature $\kappa$ of C$_3$N (103.02\,W/mK) is more than one order of
magnitude lower than that of graphene (3094.98\,W/mK).
To gain insight into the mechanisms underlying the significantly lower $\kappa$
of C$_3$N than graphene, we perform detailed mode level phonon analysis.
Due to the highly consistent phonon dispersions of C$_3$N and graphene
[Fig.~\ref{fig:dispersion}(a)], their phonon group velocities differ from each
other a little, except the FA phonon branch.
The phonon group velocity of FA for C$_3$N is much lower compared to other phonon
branches and those in graphene [Fig.~\ref{fig:mode}(a)], which is due to the
significant softness of FA.
The relatively small phonon group velocity of FA in C$_3$N is partially
responsible for its relatively smaller contribution to $\kappa$ in C$_3$N
(25.4\pct) than graphene (81.2\pct).
Considering the similar phonon group velocity and the larger specific heat
capacity of C$_3$N (18.94$\times 10^5$\,Jm$^{-3}$K$^{-1}$) than graphene
(16.19$\times 10^5$\,Jm$^{-3}$K$^{-1}$), the significantly lower $\kappa$ of
C$_3$N than graphene must stem from the smaller phonon lifetime (larger
scattering rate) based on Eq.~(\ref{eq:kappa}), which is evidently shown in
Fig.~\ref{fig:mode}(b).

It is well-known that the scattering rate is governed by two factors: the
scattering phase space and the scattering strength, which quantify how often and
how strongly the phonon mode would be scattered, respectively.
The scattering phase space is determined based on phonon dispersion only with
the criteria of energy and momentum conservation\cite{Li20141747}.
As shown in Fig.~\ref{fig:mode}(c), the scattering phase space of C$_3$N and
graphene are consistent with each other for both the absorption and emission
processes.
This is understandable in terms of the consistent phonon dispersions of C$_3$N
and graphene as shown in Fig.~\ref{fig:dispersion}(a).
Furthermore, we study the mode level Gr\"uneisen parameter that quantifies the
phonon anharmonicity and the scattering strength.
Fig.~\ref{fig:mode}(d) shows consistent results with Fig.~\ref{fig:dispersion}(b),
revealing much stronger phonon anharmonicity in C$_3$N than graphene.
The strong phonon anharmonicity in C$_3$N is consistent with the softened FA
phonon branch [Fig.~\ref{fig:dispersion}(a)] as analyzed above.
Thus, the significantly lower $\kappa$ of C$_3$N than graphene originates from
the large scattering rate [Fig.~\ref{fig:mode}(b)], which is due to the strong
phonon anharmonicity [Fig.~\ref{fig:dispersion}(b) and Fig.~\ref{fig:mode}(d)]
rather than the scattering phase space [Fig.~\ref{fig:mode}(c)].

\begin{figure}[tb]
    \centering
    \includegraphics[width=0.99\linewidth]{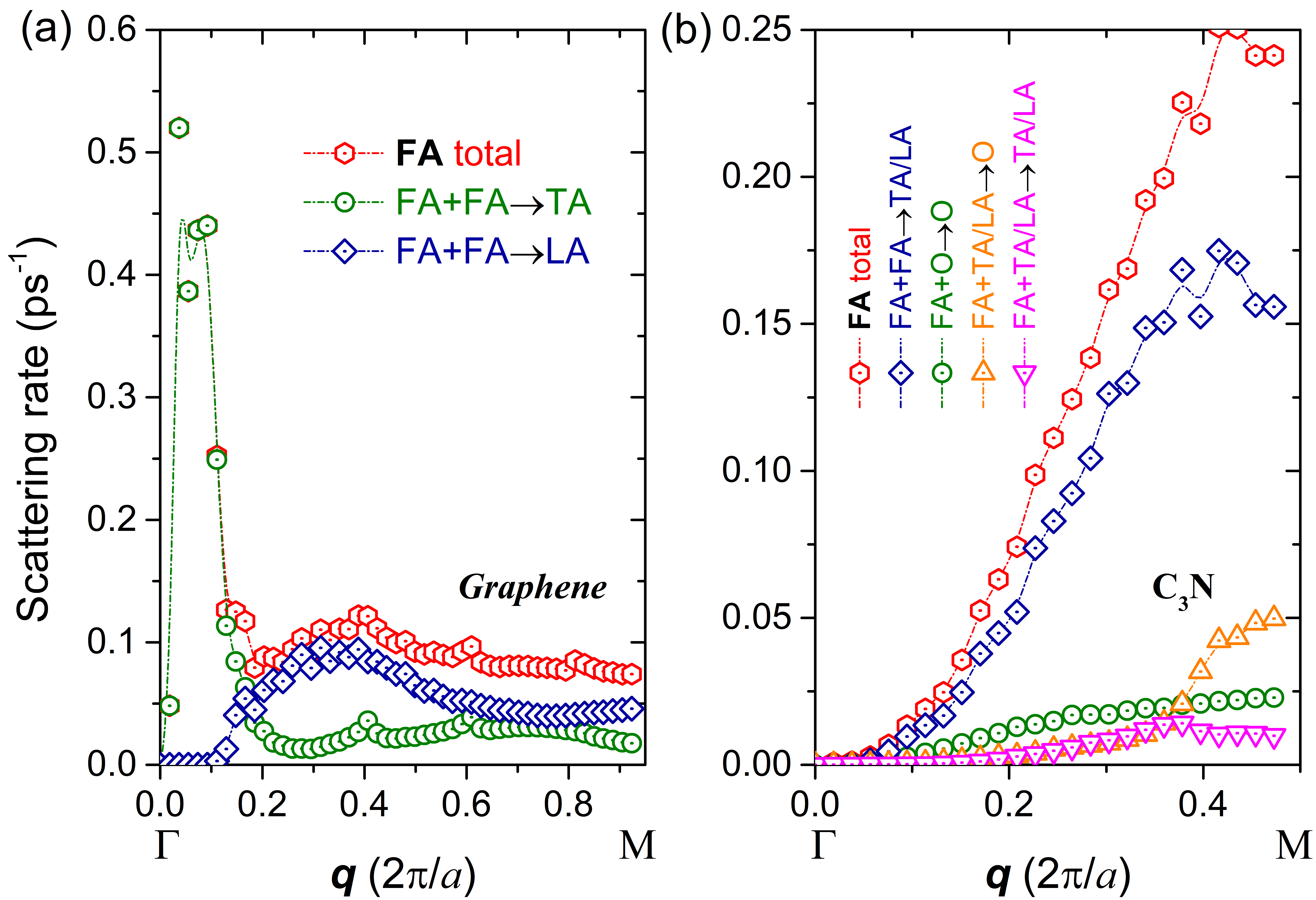}
\caption{\label{fig:FA-channels}
(Color online)
Scattering channels of FA phonon modes along the $\Gamma$-M direction for (a)
graphene and (b) C$_3$N.
There exist extra narrow scattering channels involving odd number of FA in
C$_3$N.
}
\end{figure}

As for the relatively smaller contribution from FA to $\kappa$ for C$_3$N
(25.4\pct) than graphene (81.2\pct), there are more underlying mechanisms
despite the relatively small phonon group velocity of FA in C$_3$N caused by the
significant softness of FA.
The phonon scattering channels quantifying the specific scattering process among
different phonon branches can provide fundamental insight into the phonon
scattering process, which are ruled by the conservation of energy and momentum
\cite{Phys.Rev.B.2016.93.7.075404, Nanoscale.2016.8.21.11306-11319,
NanoLett..2016.16.6.3831-3842, Phys.Rev.B.2017.95.19.195416,
Nanoscale.2017.9.12.4295-4309}.
The scattering rates for emission process are multiplied by 1/2 to avoid
counting twice for the same process.
As shown in Fig.~\ref{fig:FA-channels}(a), the scattering channels of FA phonon
branch for graphene is FA+FA$\to$TA/LA, which is governed by the so-called
symmetry-based selection rule of phonon-phonon scattering.
Due to the inversion symmetry of the planar structure of graphene, only the
scattering channels with participation of even numbers of FA are
allowed\cite{PhysRevB.82.115427}, leading to limited scattering rate of FA and
its dominating role in phonon transport.
However, it is totally different in C$_3$N where there exist also narrow
scattering channels involving odd number of FA such as FA+O$\to$O and
FA+TA/LA$\to$O/TA/LA in addition to the primary scattering channels of
FA+FA$\to$TA/LA [Fig.~\ref{fig:FA-channels}(b)].
The extra scattering channels involving odd number of FA reveal the slightly
broken inversion symmetry in C$_3$N.
The different diameter and mass of C and N atoms reduce the symmetry in C$_3$N
by changing the bond lengths or force constants, despite its similar planar
structure as graphene.
Same situation is also found in monolayer GaN\cite{Phys.Rev.B.2017.95.19.195416,
Nanoscale.2017.9.12.4295-4309}.
The extra scattering channels for FA [Fig.~\ref{fig:FA-channels}(b)] in C$_3$N
together with the relatively small phonon group velocity
[Fig.~\ref{fig:mode}(a)] lead to the lower contribution to $\kappa$ from FA
(25.4\pct) compared with that in graphene (81.2\pct).

\section{Lone-pair electrons}
Based on the above analysis, there exists strong phonon anharmonicity
[Fig.~\ref{fig:dispersion}(b) and Fig.~\ref{fig:mode}(d)] in C$_3$N,
corresponding to the softened FA phonon branch [Fig.~\ref{fig:dispersion}(a)],
which is responsible for the significantly lower $\kappa$ of C$_3$N than
graphene (Fig.~\ref{fig:kappa-T}).
Considering the similar structures and the only difference of substituting 1/4 C
atoms with N in C$_3$N compared to graphene, it is very intriguing to find the
significantly softened FA phonon branch, strong phonon anharmonicity, and low
$\kappa$ of C$_3$N.
To have a bottom-up understanding on the mechanisms underlying the low $\kappa$
of C$_3$N, we further study the strong phonon anharmonicity in C$_3$N based on
the fundamental analysis of orbital projected electronic structures.
We will show that the stereochemically active lone-pair electrons due to the
special orbital hybridization drives the remarkable phonon anharmonicity in
C$_3$N.

In graphene, the C-$s/p_x/p_y$ orbitals hybridize and contribute to the C-C
$\sigma$ bonds, while the C-$p_z$ orbital comes into being the $\pi$ bonds and
the electronic Dirac cone\cite{Balandin2012266}, which are evidently shown in
Fig.~\ref{fig:pDOS}(a).
In contrast, the bonding states in C$_3$N are totally different with an
intrinsic electronic bandgap.
The electronic bandgap is calculated to be 0.39\,eV in this study, which agrees
perfectly with previous studies and experimental
measurements\cite{AdvancedMaterials.2017.29.16.1605625, PNAS.2016.113.27.7414-7419}.
As shown in Fig.~\ref{fig:pDOS}(b), the bonding states of C atom in C$_3$N are
similar to graphene that the hybridized C-$s/p_x/p_y$ orbitals contribute to the
$\sigma$ bonds and C-$p_z$ orbital contributes to the weakened $\pi$ bonds
despite the intrinsic electronic bandgap.
The situation for the orbitals is different for the N atom where the $s$ orbital
is largely ($\sim$20\,eV) confined below the valence band, forming an isolated
band [Fig.~\ref{fig:pDOS}(c)].
As a result, the $\sigma$ bonds linking C and N atoms are jointly contributed by
the valence configuration of C-$s/p_x/p_y$ and N-$p_x/p_y/p_z$, where the $s^2$
electrons in the N-$s^2p^3$ do not participate in the bonding.
Note that the doping of $s^2$ electrons from N atoms leads to the up-shift of
Fermi level in C$_3$N compared to graphene [Fig.~\ref{fig:pDOS}(b)], which opens
the intrinsic bandgap above the maintained Dirac cone\cite{JournalofMaterialsResearch.2017.32.15.2993-3001}.

It was proposed by Petrov and Shtrum that lone-pair electrons could lead to
low $\kappa$\cite{lone-pair1962}.
The principle underlying the concept is that the overlapping wave functions of
lone-pair electrons with valence electrons from adjacent atoms induce nonlinear
electrostatic forces upon thermal agitation, leading to increased phonon
anharmonicity in the lattice and thus reducing the
$\kappa$\cite{lone-pair1962,Phys.Rev.Lett..2008.101.3.035901,Phys.Rev.Lett..2011.107.23.235901,EnergyEnviron.Sci..2013.6.2.570-578,NatPhys.2015.11.12.990-991,Angew.Chem.Int.Ed..2016.55.27.7792-7796,Phys.Rev.B.2016.94.12.125203}.
However, this is only a qualitative description.
So far, no direct evidence is available for the interactions from a fundamental
point of view.
Here, based on Figs.~\ref{fig:pDOS}(b), (c), and (e), it is clearly shown that
the non-bonding lone-pair $s$ electrons arise around N atoms in C$_3$N due to
the special orbital hybridization.
The N-$s$ electrons interact with the covalently bonding electrons of adjacent
atoms (C) due to the orbital distribution in the same energy range
[Figs.~\ref{fig:pDOS}(b,c)] and wave functions overlap [Fig.~\ref{fig:pDOS}(e)].
Additional nonlinear electrostatic force among atoms is induced by the
interactions when they thermally vibrate around the equilibrium positions.
Consequently, a more asymmetric potential energy well would be induced, which
reveals the strong phonon anharmonicity in C$_3$N [Fig.~\ref{fig:dispersion}(b)
and Fig.~\ref{fig:mode}(d)] and significantly reduces the $\kappa$
(Fig.~\ref{fig:kappa-T}).

Thus, based on the fundamental orbital hybridizations analysis of the electronic
structures, direct evidence is provided in Figs.~\ref{fig:pDOS}(b,c,e)
for the interactions between lone-pair electrons around N atoms and bonding
electrons from adjacent atoms (C).
Moreover, the microscopic picture is established to explain how the phonon
anharmonicity arises from a underlying level of electronic structure and leads
to the low $\kappa$.

\begin{figure}[tb]
    \centering
    \includegraphics[width=0.95\linewidth]{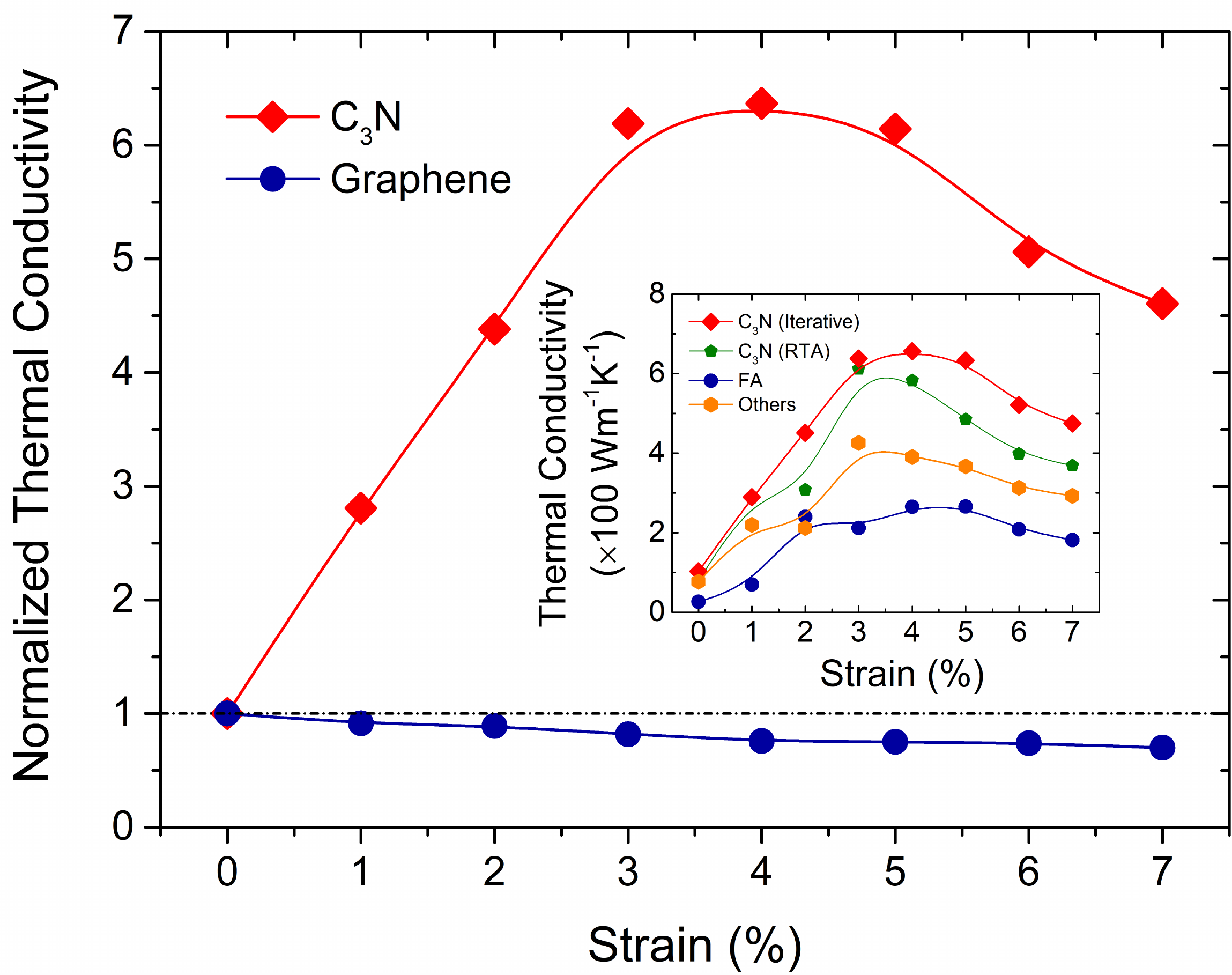}
\caption{\label{fig:kappa-strain-C3N}
(Color online)
The unusual strain enhanced $\kappa$ of C$_3$N, in sharp contrast to graphene.
All the $\kappa$ are normalized to their respective intrinsic $\kappa$ without
strain.
Inset: The absolute $\kappa$ of C$_3$N from iterative and RTA methods, together
with contributions from FA phonon branch and all other branches.
Lines are for eye guiding.
}
\end{figure}

\begin{figure*}[tb]
    \centering
    \includegraphics[width=0.99\linewidth]{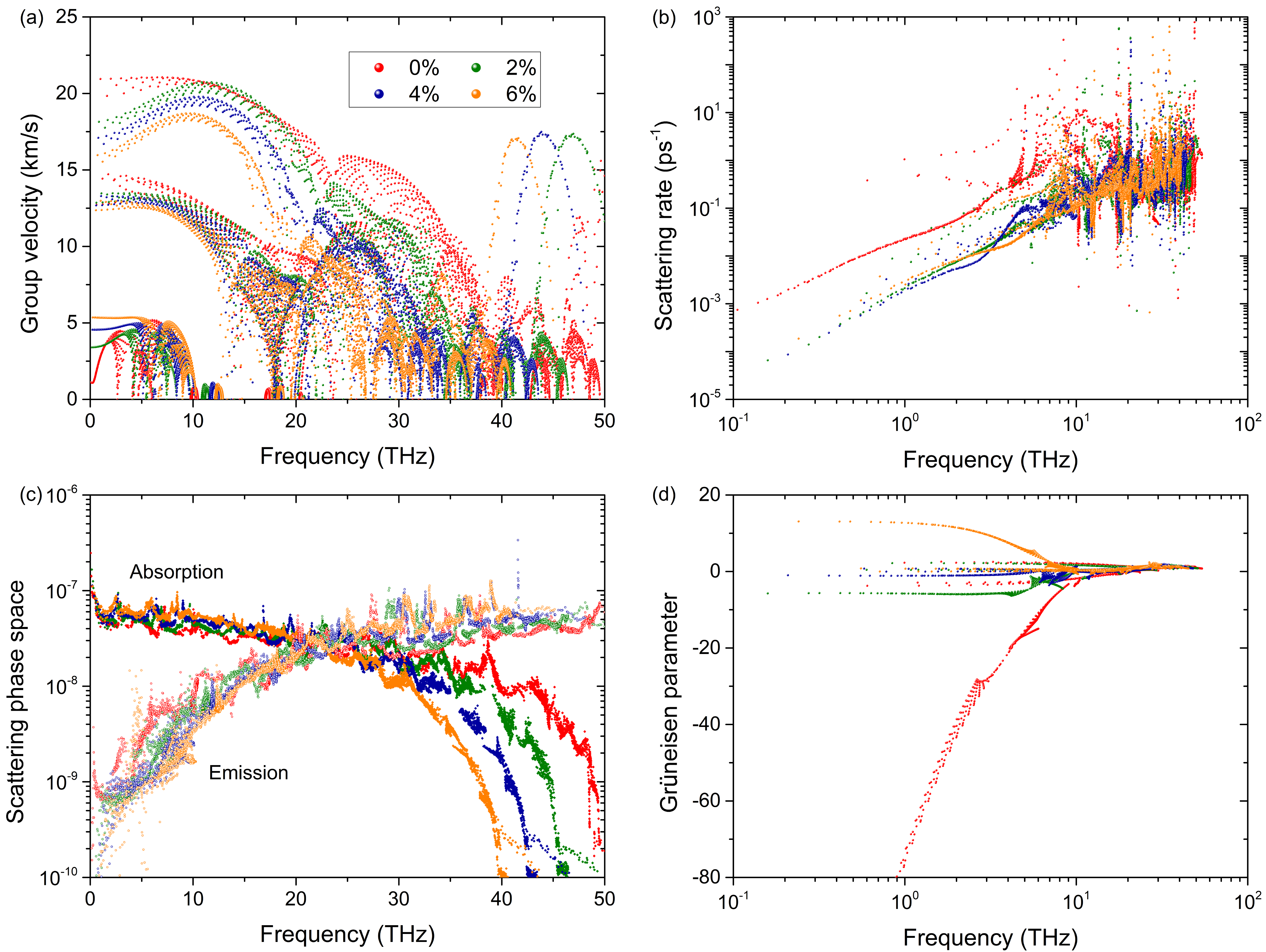}
\caption{\label{fig:strain-mode}
(Color online)
Comparison of mode level (a) phonon group velocity, (b) scattering rate, (c)
scattering phase space (absorption and emission process), and (d) Gr\"uneisen
parameter of C$_3$N between different typical strains.
}
\end{figure*}

\begin{figure}[tb]
    \centering
    \includegraphics[width=0.95\linewidth]{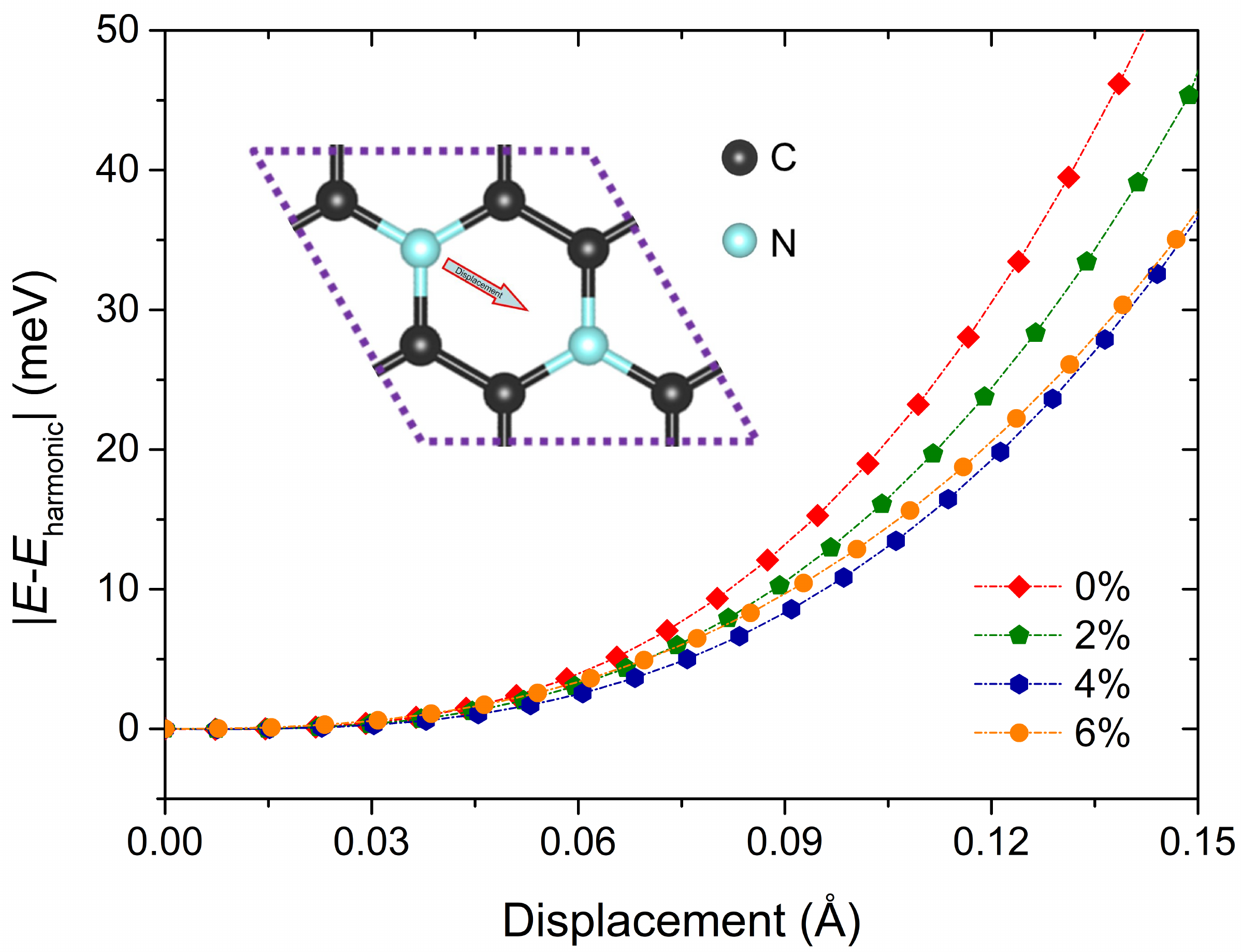}
\caption{\label{fig:anharmonicWell}
(Color online)
Energy deviation from the harmonic profile of C$_3$N with different typical
strains applied, revealing the strain modulated phonon anharmonicity.
Inset: The displacement direction of N atom in the primitive cell.
}
\end{figure}

Furthermore, it should be noted that there exists slight difference in the
electronegativity of C and N atoms.
Polarization of the C-N bonds is generated by the different electronegativity as
evidently revealed by the electron localization function (ELF), in contrast to
the nonpolarized C-C bonds [Figs.~\ref{fig:pDOS}(d) and (e)].
Consequently, the bonding electrons for the C-N bonds is relatively closer to N
atom, which contributes positively to the stronger interaction with the
non-bonding N-$s$ electrons and thus leads to a stronger phonon anharmonicity.
Furthermore, it could be expected that the phonon anharmonicity and $\kappa$ can
be effectively manipulated by altering the interaction strength.
For example, with a tensile mechanical strain applied, the interaction strength
would be weakened due to the increased distance, which is inferred to reduce the
phonon anharmonicity and thus enlarge the $\kappa$ of C$_3$N.

\section{Unusually strain enhanced thermal conductivity}

To verify the inference based on the above established microscopic picture of the
lone-pair electrons driving strong phonon anharmonicity, we further study the
effect of bilateral tensile strain on the $\kappa$ of C$_3$N, in comparison with
graphene.
Note that the C$_3$N has a much weaker stiffness (6.656\,GPa with 7\pct\ strain
applied) compared to graphene, which means that it is very easy for the strain
to be applied to C$_3$N in experiments\cite{Nanoscale.2018.10.10365}.
As shown in Fig.~\ref{fig:kappa-strain-C3N}, the $\kappa$ of graphene decreases
with tensile strain, which agrees very well with previous
reports\cite{NanoLett..2012.12.6.2673-2678,PhysRevB.89.155426,Appl.Phys.Lett..2012.101.11.111904}.
In contrast, the $\kappa$ of C$_3$N is tremendously enhanced.
It was reported in previous studies of silicene that such an enhancement by
strain engineering should be attributed to the flattening of the buckled
structure upon stretching
\cite{PhysRevB.87.195417,Phys.Rev.B.2016.93.7.075404,JournalofAppliedPhysics.2013.114.3.033526}.
Similar analysis is applied to the phosphorene which possesses
puckered structure\cite{Nanoscale.2016.8.1.483-491,J.Phys.Chem.C.2014.118.43.25272-25277,Phys.Chem.Chem.Phys..2015.17.7.4854-4858,Phys.Rev.B.2016.94.16.165445, Small.2018.14.1702465}.
In fact, the enhancement of $\kappa$ by strain engineering was thought to be
unique for 2D materials with non-planar structure\cite{Phys.Rev.B.2016.93.7.075404}.
Therefore, it is very unusual for the anomalous positive response of $\kappa$ to
tensile strains in C$_3$N, since it has a planar honeycomb structure that is
similar to graphene but different from silicene and phosphorene.

We first study the contribution to $\kappa$ of C$_3$N from different phonon
branches with strain applied.
As shown in the inset of Fig.~\ref{fig:kappa-strain-C3N}, the absolute
contribution to $\kappa$ from FA phonon branch and all the others keep
consistent with the variation trend of the total $\kappa$, and the relative
contribution keeps almost the same.
Detailed mode level analysis based on Fig.~\ref{fig:strain-mode} reveals that
the phonon group velocity and scattering phase space almost keep unchanged with
strain applied, except the little decrease of group velocity.
Note that due to phonon bunching of high-frequency phonon modes caused by the
softened phonon dispersions, the scattering phase space of absorption process
for high-frequency phonon modes decreases [Fig.~\ref{fig:strain-mode}(c)].
However, the scattering phase space of the dominating emission process for
high-frequency phonon modes keeps unchanged.
Thus, the largely strain enhanced $\kappa$ of C$_3$N is primarily due to the
overall weakened phonon-phonon scattering [Fig.~\ref{fig:strain-mode}(b)], which
is governed by the strain weakened phonon anharmonicity as quantified by the
Gr\"uneisen parameters [Fig.~\ref{fig:strain-mode}(d)].

The phonon anharmonicity can be further qualitatively characterized by the
deviation of energy potential well from the harmonic (quadratic)
profile\cite{Phys.Rev.B.2014.90.23.235201, NanoLett..2016.16.6.3831-3842}.
By displacing the N atom in C$_3$N along the bonding direction as shown in the
inset of Fig.~\ref{fig:anharmonicWell}, we calculate the energy potential well.
The energy deviation is calculated by subtracting the energy with the harmonic
profile which is obtained by fitting the first 5 points with a quadratic
function.
As shown in Fig.~\ref{fig:anharmonicWell}, the energy deviation from the
harmonic profile increases with atomic displacement increasing.
Considering the increased maximal displacement of atom with temperature
increasing, the enhanced phonon anharmonicity as revealed by the increased
energy deviation is consistent with the decreased $\kappa$ at increased
temperatures (Fig.~\ref{fig:kappa-T}).
Moreover, the deviation of energy potential well from the harmonic profile
decreases with the increased strain until the strain reaches 4\pct, and then
increases with 6\pct\ strain applied.
The strain modulated phonon anharmonicity as revealed by the energy deviation is
consistent with the strain modulated $\kappa$ (Fig.~\ref{fig:kappa-strain-C3N}),
which provides coherent understanding together with above mode level analysis
(Fig.~\ref{fig:strain-mode}).

The underlying mechanism for the anomalous strain enhanced $\kappa$ of C$_3$N
can be well understood based on the microscopic picture of the lone-pair
electrons driving phonon anharmonicity as established in this work.
With tensile strain applied, the separation distance between atoms becomes
larger.
Thus, the interaction between the lone-pair $s$ electrons around N atoms and the
bonding electrons of adjacent atoms (C) are weakened.
Consequently, the phonon anharmonicity is attenuated, reducing phonon-phonon
scattering\cite{NE.2018.50.425}.
Based on the microscopic picture, the strain enhanced $\kappa$ of C$_3$N is well
understood.
In fact, the opposite response of $\kappa$ to stretching between C$_3$N and
graphene further supports the established microscopic picture of the lone-pair
electrons driving strong phonon anharmonicity.
Furthermore, it is anticipated that other systems possessing lone-pair electrons
should also have a low $\kappa$ and their $\kappa$ can be generally enhanced by
weakening the interaction strength between the lone-pair electrons and the
bonding electrons of adjacent atoms, such as by increasing the bonding length
with tensile strain.
The possible systems possessing lone-pair electrons might be group V compounds,
just name a few, $h$-BC$_2$N, $h$-BN, $h$-AlN, $h$-GaN, and $h$-BAs.

\section{Conclusions}

In summary, we have performed a comparative study of phonon thermal transport
between monolayer C$_3$N and graphene.
The $\kappa$ of C$_3$N shows an anomalous temperature dependence, which is
totally different from that for common crystalline materials and deviates
largely from the well-known $\kappa\sim 1/T$ relationship.
Consequently, the $\kappa$ of C$_3$N at high temperatures is larger than the
expected value that follows the general trend of $\kappa \sim 1/T$, which would
be much beneficial for the applications in nano- and opto-electronics in terms
of efficient heat dissipation.
Moreover, it is very intriguing to find that the $\kappa$ of C$_3$N is
substantially lower than graphene, considering the similar structures and the
only difference of substituting 1/4 C with N atoms in C$_3$N compared to
graphene.
The large scattering rate is responsible for the significantly low $\kappa$ of
C$_3$N, which is due to the strong phonon anharmonicity.
By deeply analyzing the orbital projected electronic structure, we establish a
microscopic picture of the lone-pair electrons driving strong phonon
anharmonicity.
Direct evidence is provided for the interactions between lone-pair electrons
(N-$s$) and bonding electrons from adjacent atoms (C), which induce nonlinear
electrostatic force among atoms when they thermally vibrate around the
equilibrium positions, leading to the strong phonon anharmonicity and
significantly low $\kappa$ of C$_3$N.
Furthermore, the $\kappa$ of C$_3$N is unexpectedly enlarged by applying
bilateral tensile strain despite the planar honeycomb structure of C$_3$N
(similar to graphene, with no buckling or puckering), which is in sharp contrast
to the strain induced $\kappa$ reduction in graphene.
The anomalous positive response of $\kappa$ to tensile strain is attributed to
the attenuated interaction between the lone-pair $s$ electrons around N atoms
and the bonding electrons of neighboring C atoms, which reduces phonon
anharmonicity.
The opposite response of $\kappa$ to mechanical strain between C$_3$N and
graphene further supports the established microscopic picture of the lone-pair
electrons driving strong phonon anharmonicity.
We propose that other systems possessing lone-pair electrons would also have low
$\kappa$ and the $\kappa$ can be generally enhanced by weakening the interaction
strength between the lone-pair electrons and the bonding electrons of adjacent
atoms, such as by increasing the bonding length with tensile mechanical strain.
The microscopic picture for the lone-pair electrons driving phonon anharmonicity
established from the fundamental level of electronic structure deepens our
understanding of phonon transport in 2D materials and would also have great
impact on future research in micro-/nano-scale thermal transport such as
materials design with targeted thermal transport properties.

\begin{acknowledgments}
Simulations were performed with computing resources granted by RWTH Aachen
University under projects jara0168 and rwth0223.
This work is supported by the Deutsche Forschungsgemeinschaft (DFG) (Project
number: HU 2269/2-1).
M.H.\ acknowledges the start-up fund from the University of South Carolina.
Z.Q.\ is supported by the National Natural Science Foundation of China (Grant
No.\ 11847158) and the China Postdoctoral Science Foundation (2018M642776).
Research reported in this publication was supported in part by the NSF and SC
EPSCoR/IDeA Program under award number (NSF Award Number OIA-1655740 and SC
EPSCoR/IDeA Award Number 19-SA06).
\end{acknowledgments}

\bibliography{bibliography}

\end{document}